\documentclass[preprint]{elsarticle}
\usepackage{amsmath}
\usepackage{xcolor}
\usepackage{hyperref}
\usepackage{ulem}

\begin{document}

\begin{frontmatter}
  \title{Efficient Fourier Basis Particle Simulation}
  \author[cips]{Matthew S. Mitchell}
  \author[cips]{Matthew T. Miecnikowski}
  \author[appm]{Gregory Beylkin}
  \author[cips]{Scott E. Parker}

  \address[cips]{Department of Physics, University of Colorado, Boulder, CO 80303 USA}
  \address[appm]{Department of Applied Mathematics, University of Colorado,
    Boulder, CO 80303 USA}

  \begin{abstract}
    The standard particle-in-cell algorithm suffers from grid heating. There exists a gridless
    alternative which bypasses the deposition step and calculates each Fourier mode of the charge
    density directly from the particle positions. We show that a gridless method can be computed
    efficiently through the use of an Unequally Spaced Fast Fourier Transform (USFFT) algorithm.
    After a spectral field solve, the forces on the particles are calculated via the inverse USFFT
    (a rapid solution of an approximate linear system) \cite{beylkin_fast_1995,dutt_fast_1993}. We
    provide one and two dimensional implementations of this algorithm with an asymptotic runtime of
    $O(N_p + N_m^D \log N_m^D)$ for each iteration, identical to the standard PIC algorithm (where
    $N_p$ is the number of particles and $N_m$ is the number of Fourier modes, and $D$ is the
    spatial dimensionality of the problem) We demonstrate superior energy conservation and reduced
    noise, as well as convergence of the energy conservation at small time steps.
  \end{abstract}
\end{frontmatter}

\section{Introduction}

A common approach to numerically solving the Vlasov-Poisson system is to represent
the distribution function using particles, with fields solved on a grid and interpolated at the
particle positions \cite{birdsall_plasma_1985,hockney_computer_1988}. This scheme, known as the
particle-in-cell (PIC) method, has been enormously successful for simulating plasmas and is used in
a wide variety of applications, but does not conserve total energy \cite{langdon_effects_1970}.
Energy-conserving schemes based on variational formulations have been proposed
\cite{lewis_energy-conserving_1970,eastwood_virtual_1991}, but generally do not conserve momentum
because of a lack of translational invariance \cite{evstatiev_variational_2013}, though the momentum
error can be kept small in many cases due to the choice of integrator \cite{chen_energy-_2011}.

In addition to the lack of energy conservation, PIC also suffers from a finite grid instability, in
which sufficiently high Fourier modes experience exponential growth
\cite{okuda_nonphysical_1972,lindman_dispersion_1970} due to coupling with lower modes. This
phenomenon causes numerical heating, which saturates when the Debye length is on the order of the
grid spacing \cite{okuda_nonphysical_1972,evstatiev_variational_2013}. The finite grid instability
is of particular relevance to problems involving cold plasmas or multiple length scales
\cite{okuda_nonphysical_1972,brackbill_method_1994}. Various approaches have been proposed to reduce
this instability, such as the use of smoother particle shapes \cite{jacobs_high-order_2006}, grid
jiggling \cite{brackbill_method_1994}, filtering \cite{hockney_quiet_1974}, and high-order Galerkin
methods \cite{jacobs_high-order_2006}, but many of these schemes suffer from issues such as a lack
of charge conservation and high computational costs.

Several energy-conserving particle algorithms
\cite{lewis_energy-conserving_1970,eastwood_virtual_1991} based on the Lagrangian formulated by Low
\cite{low_lagrangian_1958} have been suggested as alternatives to PIC. Charge-conserving approaches
\cite{chen_energy-_2011,markidis_energy_2011} using implicit time integration have also been
proposed. Energy-conserving algorithms have the benefit of eliminating numerical heating, but suffer
from several drawbacks. They generally do not conserve momentum
\cite{langdon_energy-conserving_1973}, sometimes suffer from increased noise, and heavily restrict
the choice of particle shape \cite{evstatiev_variational_2013}. However, these methods have seen
widespread use due to their efficiency and simplicity.

It has been demonstrated that exact energy and momentum conservation in a particle code can be
achieved by depositing charge on a truncated Fourier basis \cite{evstatiev_variational_2013}. This
method has also been shown to eliminate the finite grid instability and reduce coupling between
modes \cite{huang_finite_2016}. However, due to the poor scaling and high computational cost of this
approach ($O(N_m N_p)$, where $N_m$ is the number of Fourier modes and $N_p$ is the number of
particles), it has not been seriously considered in practice.

We present a similar algorithm to the one proposed in \cite{evstatiev_variational_2013}, in which we
model the charge density as a sum of shape functions in continuous space and perform the field solve
with a truncated Fourier series. However, we reduce the computation time to $O(N_p + N_m \log N_m)$,
which is equivalent to that of conventional PIC with a spectral field solve, by making use of an
Unequally Spaced FFT (USFFT) \cite{beylkin_fast_1995,dutt_fast_1993}.

This paper is organized as follows: we begin by reviewing the standard PIC method in Section
\ref{sec:methods}. We then propose a gridless algorithm based on \cite{evstatiev_variational_2013}
and demonstrate that it can be made efficient via the USFFT. In Section \ref{sec:results}, we
analyze the results of several numerical experiments, comparing our code to an implementation of
conventional PIC and providing experimental confirmation of its energy conservation. A proof of
energy conservation in the continuous-time limit, as well as momentum conservation, is given in
Section \ref{sec:cons}. In Section \ref{sec:perf}, we present an analysis of the algorithm's
asymptotic time complexity (in both one and two dimensions) and a possible parallelization scheme
along with the results of several scaling experiments. Section \ref{sec:concl} contains a summary of
our results and a discussion of possible generalizations.

\section{Numerical Methods} \label{sec:methods}

\newcommand{\pder}[2]{\frac{\partial #1}{\partial #2}}
\newcommand{\vb}[1]{\mathbf{#1}}

For a collisionless electrostatic plasma, the time evolution of the distribution function
$f(\vb{x}, \vb{v})$ is given by the Vlasov-Poisson system:

\begin{equation} \label{eq:vlas1}
  \pder{f}{\vb{t}}
  + \vb{v} \cdot \pder{f}{\vb{x}}
  + \frac{q}{m} \vb{E} \cdot \pder{f}{\vb{v}} = 0,
\end{equation}

\begin{equation} \label{eq:vlas2}
  \nabla \cdot \vb{E} = \frac{\rho}{\epsilon_0}.
\end{equation}

\subsection{Particle-in-Cell Method} \label{subsec:picrev}

The Vlasov-Poisson system is commonly solved via the particle-in-cell (PIC) algorithm
\cite{birdsall_plasma_1985,hockney_computer_1988}, which tracks particles in continuous phase space,
while fields are tracked on a spatial grid with $N_g$ points. Each timestep begins by calculating
the charge density $\rho(x)$ at the grid points from the particle positions (denoted $\vb{X}_i$).
This is accomplished via a deposition of charge on the grid, such that

\begin{equation} \label{eq:picdep}
  \rho_\vb{x} = \sum_i q S(\vb{X}_i - \vb{x})
\end{equation}

where $S$ is some shape function corresponding to the weighting scheme. $E$ and $\phi$ are then
calculated at the grid points from $\rho$, usually via spectral methods, as follows:

\begin{align}
  \tilde{\rho}_\vb{k} &= \sum_{\vb{x}} \rho_\vb{x} e^{-i \vb{k} \cdot \vb{x}},    \label{eq:picslv1}\\
  \tilde{\vb{E}}_\vb{k} &= \frac{1}{i \vb{k} \epsilon_0} \tilde{\rho}_\vb{k},     \label{eq:picslv2}\\
  \tilde{\phi}_\vb{k} &= \frac{1}{k^2\epsilon_0} \tilde{\rho}_\vb{k},             \label{eq:picslv3}\\
  \vb{E}_\vb{x} &= \sum_{\vb{k}} \tilde{\vb{E}}_\vb{k} e^{i \vb{k} \cdot \vb{x}}. \label{eq:picslv4}
\end{align}

After the field solve, the forces on the particles are calculated by interpolating the E-field from
the grid to the particle positions as

\begin{equation} \label{eq:picint}
  \vb{F}_i = \sum_\vb{x} q \vb{E}_\vb{x} S(\vb{X}_i - \vb{x}).
\end{equation}

In this paper, we assume without loss of generality that the same weighting scheme is used for both
deposition and interpolation, but this need not be the case.

The particles are then pushed forward, necessitating discretization in time. Our implementation of
PIC (as well as our algorithm, which we discuss in section \ref{subsec:algo}) makes use of the
leapfrog integrator. Position and velocity are tracked at alternating timesteps, as follows:

\begin{align} \label{eq:leap}
  \vb{X}_i^{n+1} &= \vb{X}_i^{n} + \vb{v}_i^{n + 1/2} \Delta t\\
  \vb{v}_i^{n+1/2} &= \vb{v}_i^{n-1/2} + \frac{\vb{F}_i^{n}}{m} \Delta t .
\end{align}

\subsection{Particle-in-Fourier Method} \label{subsec:algo}

Our algorithm, termed Particle-in-Fourier (or PIF), begins by following established gridless
algorithms \cite{evstatiev_variational_2013}. Instead of depositing the charge on a grid, we treat
$\rho(\vb{x})$ as a sum of shape functions in continuous space and ``deposit'' on a truncated
Fourier basis. This can be accomplished by calculating each mode of $\tilde{\rho}$ directly from the
particle positions, as

\begin{align}\label{eq:ftdep}
    \tilde{\rho}_\vb{k} &= \int_0^L d\vb{x}\; e^{-i \vb{k} \cdot \vb{x}} \rho(\vb{x}),\\
                        &= \int_0^L d\vb{x}\; e^{-i \vb{k} \cdot \vb{x}}
                          \sum_{i=1}^{N_p} q S(\vb{x} - \vb{X}_i),\\
    &= q \tilde{S}_\vb{k} \sum_{i=1}^{N_p} e^{-i \vb{k} \vb{X}_i}, \label{eq:ftdep3}
\end{align}

where $\vb{k}$ is of the form
$2 \pi (\frac{n_x}{L_x} \vb{\hat{x}} + \frac{n_y}{L_y} \vb{\hat{y}} + \frac{n_z}{L_z}
\vb{\hat{z}})$, for integers $n_x, n_y, n_z \in [-\frac{N_m}{2},\frac{N_m}{2}]$, $N_m$ being the
number of modes (analogous to the number of grid cells in PIC). $\tilde{S}_\vb{k}$ denotes the Fourier
transform of the shape function, given by

\begin{equation} \label{eq:sk}
  \tilde{S}_\vb{k} = \int_0^L d\vb{x}\; e^{-i\vb{k} \cdot \vb{x}} S(\vb{x}).
\end{equation}

A unique feature of the PIF method is the ability to use a physical particle shape given by
$S(\vb{x})$ as an alternative to filtering. A natural choice is a Gaussian particle shape with a
spatial width chosen as a filtering parameter. Alternatively, a physical delta-function particle
shape can be used without any mathematical complication. It is important to note that there is no
computational penalty for using higher order shape functions with PIF. Because convolution with the
shape function is simply a multiplication in Fourier space and $\tilde{S}_\vb{k}$ is typically
evaluated analytically (and the results can be reused between timesteps), the step time is
independent of the weighting scheme. This permits the use of arbitrary shape functions with no
additional computational cost.

We now perform the field solve in a manner identical to conventional PIC, finding
$\tilde{\vb{E}}_\vb{k}$ and $\tilde{\phi}_\vb{k}$ from Eq.\ (\ref{eq:picslv2}) and
(\ref{eq:picslv3}). Then, using a similar approach to the deposition, we can determine the forces on
the particles by summing over all modes of $\vb{E}$:

\begin{align}\label{eq:ftforce}
  \vb{F}_i &= \int_0^L d\vb{x}\; q \vb{E}(\vb{x}) S(\vb{x} - \vb{X}_i), \\
           &= \sum_{\vb{k}} e^{i \vb{k} \cdot \vb{X}_i} q \tilde{\vb{E}}_\vb{k} \tilde{S}_\vb{k}.
             \label{eq:ftforce2}
\end{align}

\subsection{Unequally Spaced Fast Fourier Transform (USFFT)} \label{sec:uf}

Our algorithm relies on rapid evaluation of two computationally expensive sums:
\begin{equation}
  \sum_i e^{-i \vb{k} \cdot \vb{X}_i},
\end{equation}
which must be performed for every mode, and
\begin{equation}
\sum_\vb{k} \tilde{S}_\vb{k} \tilde{\vb{E}}_\vb{k} e^{i \vb{k} \cdot \vb{X}_i},
\end{equation}
which must be performed for every particle. Naively, computing these sums requires $O(N_p N_m^D)$
operations for $N_p$ particles and $N_m$ modes in each direction. However, we can reduce the
computational cost by using an Unequally Spaced FFT (USFFT) \cite{beylkin_fast_1995,
  dutt_fast_1993}.

Let us briefly describe the basics of USFFT algorithm (in one dimension, for simplicity). We want to
  compute the values of the Fourier transform of a generalized function

\begin{equation}
  f\left(x\right)=\sum c_{m}\delta\left(x-x_{m}\right)
\end{equation}

that is,

\begin{equation}
  \widehat{f}\left(\xi\right)=\sum_{m}c_{m}e^{-i\xi x_{m}},
\end{equation}

in an interval $\left|\xi\right|\le c$. A generic approach is to replace $f$ with a smooth (i.e.
many times differentiable) periodic function $g$, such that its Fourier transform accurately
approximates the Fourier transform $\widehat{f}$ in the interval $\left|\xi\right|\le c$. Once the
function $g$ and its values are available, we can then use the FFT to evaluate the Fourier transform
$\widehat{g}$ (instead of $\widehat{f}$) at an equally spaced grid. Note that, due to the required
smoothness of $g$, $\widehat{g}$ must decay rapidly.

If we were to have direct access to the Fourier domain, then it would easy to construct $g$ by simply
multiplying $\widehat{f}$ by a smooth function $\widehat{w}$ (the so-called window function), such
that it is close to $1$ for $\left|\xi\right|\le c$ and rapidly decays to zero for
$\left|\xi\right|\le c$. Multiplying by $\widehat{w}$ is equivalent to the convolution with $w$ in
space, thus leading to a relatively fast algorithm suggested in \cite{press_fast_1989}. However, if
the transition of $\widehat{w}$ from one to zero is rapid, then the convolution with $w$ will be
computationally expensive; if the transition is gradual, then the interval in the Fourier domain
becomes large and, again, the computational cost is significant. An algorithmic solution is to
construct the window $\widehat{w}$ as a ratio of two functions, a rapidly decaying numerator that is
applied in space as a convolution, and a denominator, that is applied in the Fourier domain as a
``compensating'' factor.

There are many possible choices for the numerator of $\widehat{w}$. \cite{dutt_fast_1993} uses a
Gaussian (however, the accuracy estimates in this paper are overly pessimistic), while the USFFT
uses b-splines \cite{beylkin_fast_1995}. The gain in speed using this approach is significant and it
is used in higher dimensions in a similar manner. Importantly, the error is controlled (see
\cite{beylkin_fast_1995}) so that the algorithm can be used as a ``black box.''

In one dimension, the USFFT algorithm computes the sum

\begin{equation} \label{eq:uff}
  \tilde{g}_n = \sum_{l=1}^{N_p} g_l e^{\pm 2 \pi i x_l n}
\end{equation}
for all $-N_m/2 \le n \le N_m/2$, where $x_l \in [-1/2,1/2]$ (equivalent to $x_l \in [0,1]$)
and $g_l$ are arbitrary complex coefficients. The dual version of the USFFT rapidly evaluates the
sum

\begin{equation} \label{eq:ufd}
  g(x_l) = \sum_{n=-N_m/2}^{N_m/2-1} \tilde{g}_n e^{\pm 2 \pi i x_l n},
\end{equation}
where $\tilde{g}_n$ are arbitrary complex coefficients.

The computational cost of the USFFT is $O(N_p + N_m \log N_m)$ in one dimension, and
$O(N_p + N_m^D \log N_m^D)$ for the higher-dimensional implementations. It is important that the
accuracy of the USFFT is user-controlled, and since the accuracy is guaranteed by the algorithm, one
can use the USFFT in a manner similar to a conventional FFT (tight accuracy estimates can be found
in \cite{beylkin_fast_1995}).

The USFFT algorithm uses the FFT as one of the steps, and for this reason, it is helpful to estimate
the computational cost of the USFFT in terms of that of the FFT. The double precision USFFT in one
dimension is approximately $2.5-3$ times slower than the FFT of the same size. In two dimensions,
the factor is approximately $10-20$.

We use the USFFT by setting $g_l = 1$ and $x_l = X_l / L$ for all particles in 1D, so that we can
efficiently compute the sum in Eq.\ (\ref{eq:ftdep3}). Similarly, by taking $\tilde{g}_i$ as the
amplitude of each mode of the electric field, we can compute the sum in Eq.\ (\ref{eq:ftforce2}) (in
two and three dimensions, these must be done separately for each component of the electric field).
As a result, PIF has the same computational complexity as standard PIC. We analyze this further in
Section \ref{sec:perf}.

\section{Numerical Tests} \label{sec:results}

In this section, we test the behavior of the two methods on two classic kinetic problems in plasma
physics for which PIC has been well verified with theory, namely the two stream instability and
linear Landau damping. Results for the two stream instability presented in Sec. 3.1 are using a 1D
implementation. Results for Landau damping are presented for both 1D and 2D implementations in Sec.
3.2.

\subsection{Two Stream Instability} \label{subsec:2stream}

An implementation of PIF in one and two dimensions was created, using a serial implementation of the
USFFT \cite{beylkin_fast_1995}. The performance and energy conservation of the code were compared to
an implementation of conventional 1D and 2D  PIC with a linear particle shape,
first order interpolation, and a spectral field solve. Shared memory parallelism was implemented in
both codes using OpenMP (see Section \ref{sec:perf} for more details). Source code can be found at
\url{https://github.com/matt2718/ftpic}.

\begin{figure}[h]
  \centering
  \includegraphics[width=8cm]{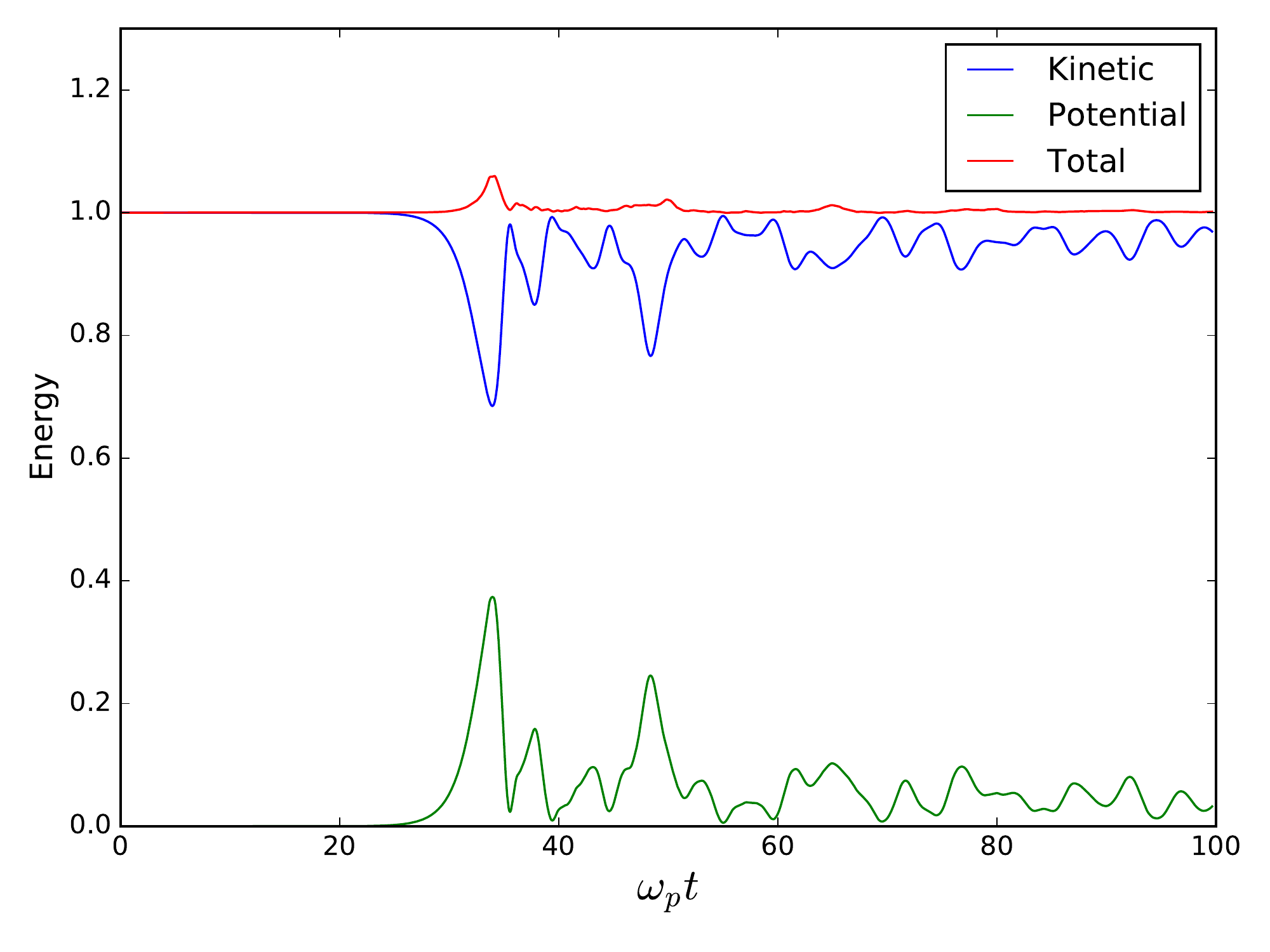}
  \caption{Kinetic, potential, and total energy for the PIC method applied to the 1D two stream
    instability. Total energy is not conserved.}
  \label{fig:2streamo}
\end{figure}

\begin{figure}[h]
  \centering
  \includegraphics[width=8cm]{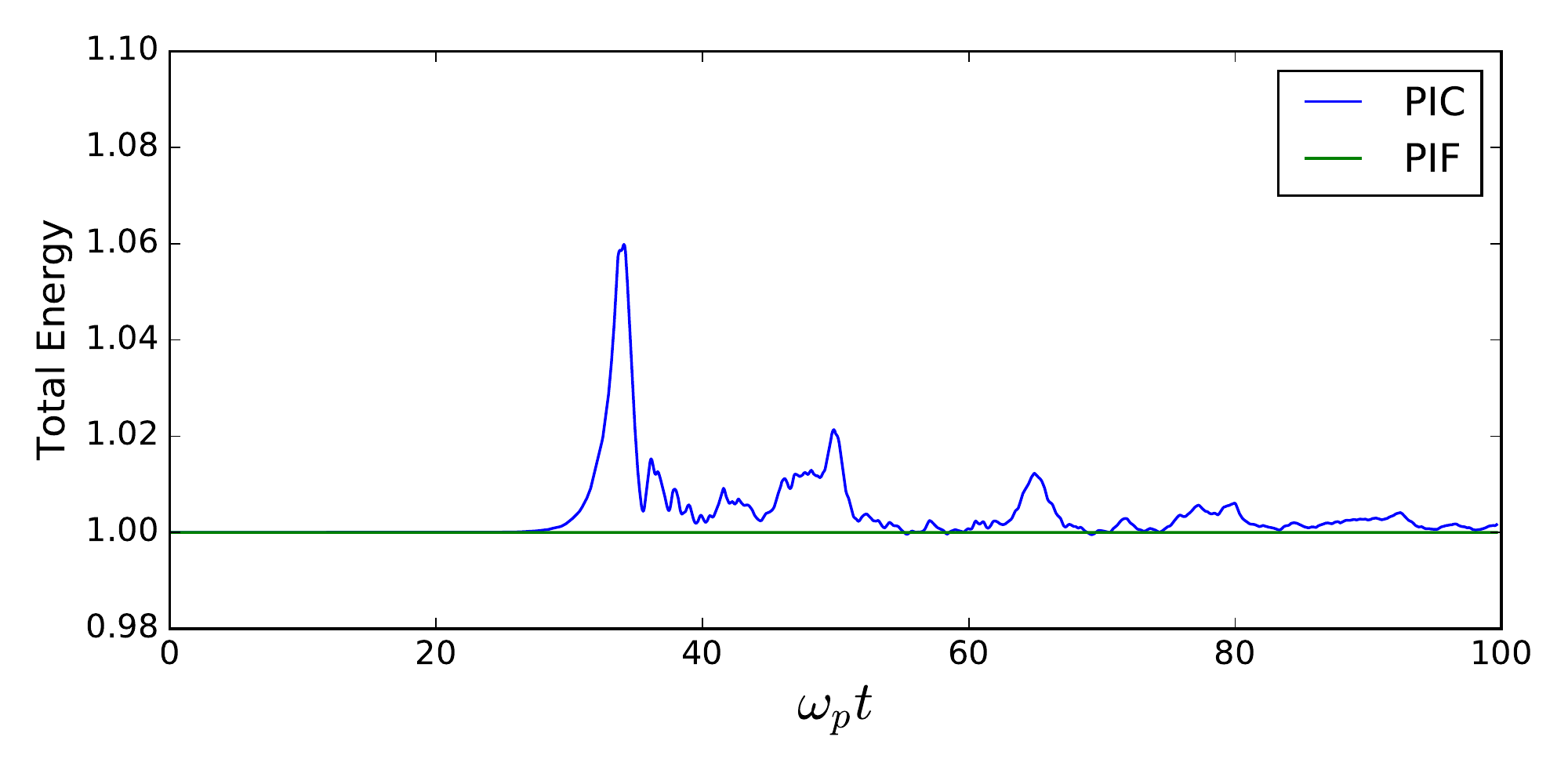}
  \caption{Comparison of total energy over time in 1D PIF and PIC simulations of the two stream
    instability, demonstrating the superior energy conservation of PIF. For PIF, the maximum
      deviation from the initial energy was $4 \times 10^{-6}$.}
  \label{fig:ecomp}
\end{figure}

The superior energy conservation of PIF becomes evident when we examine the the two stream
instability. The 1D implementations of PIF and PIC were used to simulate the mixing of two
counter-moving electron beams against a neutralizing background. Both codes were run with 10000
particles and 16 modes/grid cells for 20000 time steps, with the effective grid spacing equal to
approximately $0.88 \lambda_D$. Fig.\ \ref{fig:2streamo} plots the normalized field, kinetic, and
total energy of each system. It is clear that the total energy is not conserved for conventional
PIC. Fig.\ \ref{fig:ecomp} illustrates the total energy over time for both simulations,
demonstrating energy conservation in PIF, but not in PIC. The peak in the energy error shown in Fig.
\ref{fig:ecomp} happens at nonlinear saturation. Although the number of grid cells used is lower
than would be seen in practice, this was chosen to simply illustrate a specific result---that PIF
conserves energy \textit{regardless of the number of modes used}, and requires fewer modes to
reproduce physical behavior.

\begin{figure}[h]
  \centering
  \includegraphics[width=0.9\textwidth]{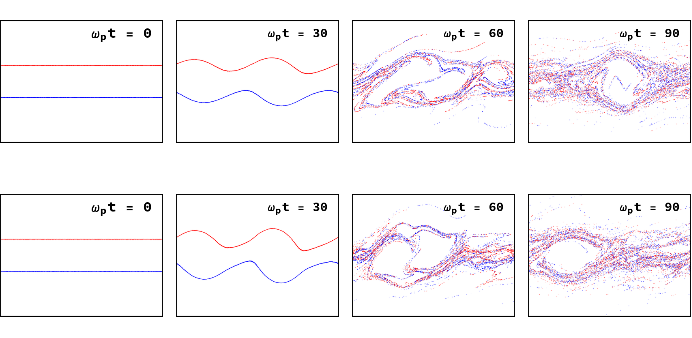}
  \caption{Time evolution of phase space for PIC (top) and PIF (bottom) simulations of a two stream
    instability. The beams are represented by different colors to demonstrate mixing, but all
    particles are identical.}
  \label{fig:phase}
\end{figure}

Aside from the difference in energy conservation, the two codes produced similar results. Fig.\
\ref{fig:phase} illustrates the evolution of phase space over time in the PIF and PIC
simulations, which we see produce qualitatively similar results. The two beams are
represented by different colors, but all particles are identical. Furthermore, momentum was
  conserved within both the PIF and PIC simulations to within floating point error.

\subsection{Landau Damping} \label{subsec:landau}

\begin{figure}[h]
    \centering
    \begin{minipage}{0.48\textwidth}
        \centering
        \includegraphics[width=1\textwidth]{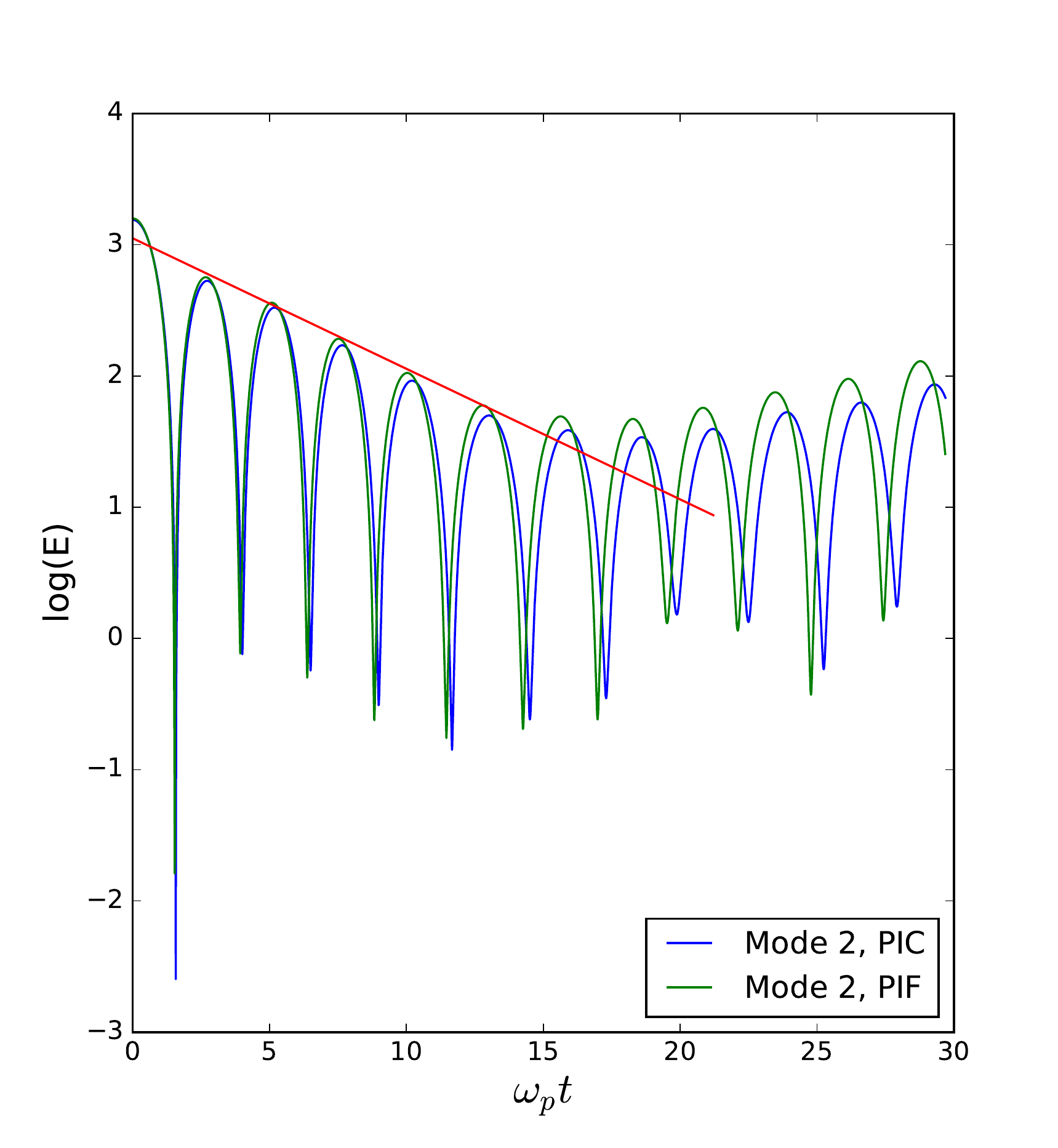}
        \caption{Landau damping of the 2nd Fourier mode in a 32-cell simulations. We see
          agreement between the theoretical rate and the rate produced by PIF, but not that of PIC.}
        \label{fig:landau32}
    \end{minipage}\hfill
    \begin{minipage}{0.48\textwidth}
        \centering
        \includegraphics[width=1\textwidth]{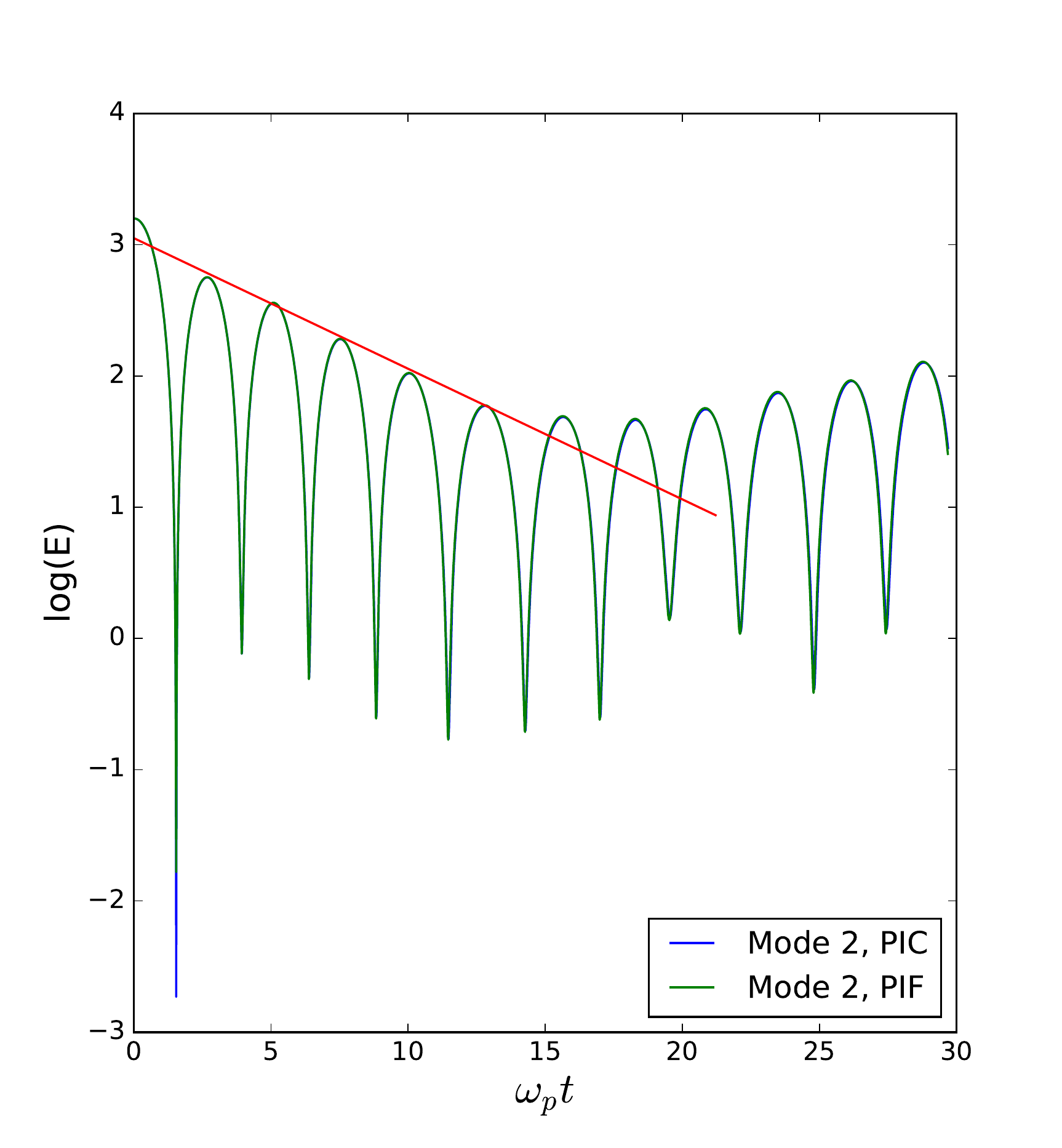}
        \caption{Landau damping of the 2nd Fourier mode in a 128-cell simulations. The results of
          both algorithms agree with the damping rate predicted by theory.}
        \label{fig:landau128}
    \end{minipage}
\end{figure}

Our 1D implementations of PIF and conventional PIC were used to simulate a Maxwellian plasma of 10000
 electrons with a neutralizing ion background, with a sinusoidal density
perturbation of the second Fourier mode given by $n_1 = a \cos(4 \pi x / L)$. This was performed for
both 32 grid cells in PIC and 32 modes in PIF ($\Delta x = 1.0 \lambda_D$) and for 128 grid cells in
PIC and 128 modes in PIF ($\Delta x = 0.25 \lambda_D$). The observed damping rate of the second mode
of the E-field was compared to the theoretical rate, illustrated in Fig.\ \ref{fig:landau32} and
Fig.\ \ref{fig:landau128}.

At small grid spacings, we see that the results of the PIC simulation converge to those of PIF. PIF,
however, accurately reproduces the theoretical damping rate regardless of the number of modes
present. This is a somewhat contrived example (in practice, the grid spacing would not be equal to
the Debye length), but it demonstrates that PIF can accurately reproduce physical behavior even when
few modes are kept.

\begin{figure}[h]
  \centering
  \includegraphics[width=8cm]{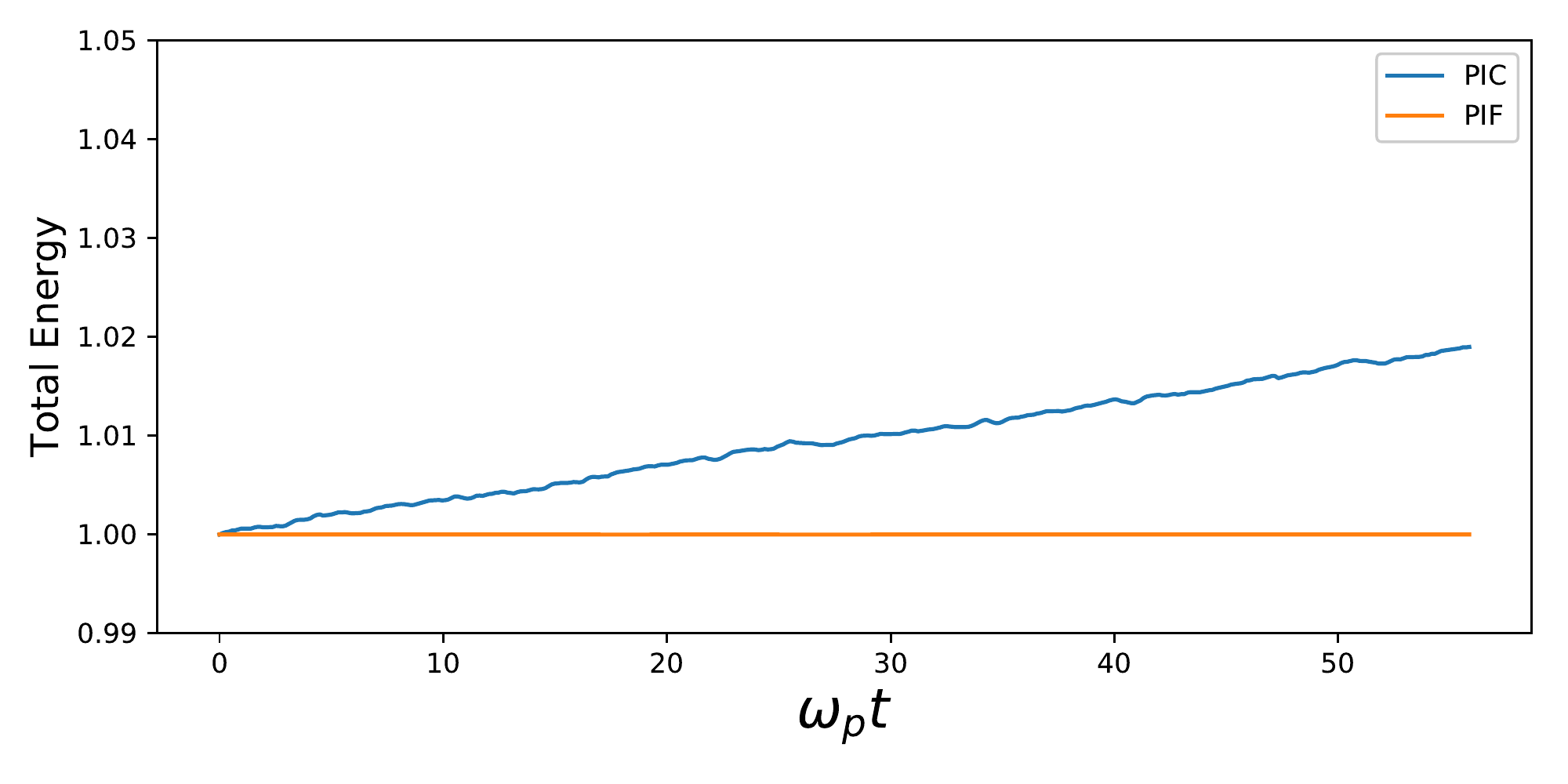}
  \caption{Comparison of total energy over time in 2D PIF and PIC simulations of Landau damping,
    demonstrating the superior energy conservation of PIF in two dimensions.}
  \label{fig:lcomp}
\end{figure}

A similar test case was run in two dimensions with 10000 particles, 128 grid cells/modes, and
$\Delta x = 0.25 \lambda_D$. Similar behavior was observed, and significant grid heating was found
in the two dimensional version of PIC, while PIF demonstrated perfect energy conservation (Fig.
\ref{fig:lcomp}).

\section{Conservation Properties and Numerical Error} \label{sec:cons}

It is well-established that the standard PIC algorithm does not conserve energy; a full proof is
given by Langdon \cite{langdon_effects_1970}. We can prove that PIF conserves energy in the
continuous-time limit by taking the field and kinetic energies to be

\begin{align}
  U_E &= \frac{1}{2} \sum_k \tilde{\rho}_k \tilde{\phi}_k, \label{eq:endef1}\\
  T &= \sum_i \frac{1}{2} m \vb{v}_i^2, \label{eq:endef2}
\end{align}

and expanding their time derivatives as

\begin{align}
  \frac{d}{dt} U_E &= \frac{d}{dt} \left(\frac{1}{2}
                     \sum_k \tilde{\rho}_k \frac{\tilde{\rho}_k}{\epsilon_0 k^2} \right) \\
                   &= \sum_k \frac{1}{\epsilon_0 k^2}
                     \tilde{\rho}_k \dot{\tilde{\rho}}_k \\
                   &= \sum_k \frac{1}{\epsilon_0 k^2} \tilde{\rho}_k
                     \frac{d}{dt} \left( \sum_i q \tilde{S}_k e^{-i k X_i} \right) \\
                   &= -\sum_i \sum_k \frac{q \vb{v}_i}{i k \epsilon_0}
                     \tilde{S}_k \tilde{\rho}_k  e^{-i k X_i}, \label{eq:uder}
\end{align}

and

\begin{align}
  \frac{d}{dt} T &= \sum_i m \dot{\vb{v}}_i \vb{v}_i \\
                 &= \sum_i \left( \sum_k q \tilde{S}_k \tilde{E}_k e^{i k X_i} \right) \vb{v}_i \\
                 &= \sum_i \sum_k \frac{q \vb{v}_i}{i k \epsilon_0} \tilde{S}_k \tilde{\rho}_k e^{i
                   k X_i}. \label{eq:tder}
\end{align}

Eq. (\ref{eq:tder}) is the negative complex conjugate of Eq. (\ref{eq:uder}). Because the field
quantities are Hermitian in Fourier space, and the sum in $k$ runs from $-\frac{2\pi}{L} N_m$ to
$\frac{2\pi}{L} N_m$ (and because both the kinetic and field energies must be real), Eqs.
(\ref{eq:uder}) and (\ref{eq:tder}) must sum to 0, so total energy is conserved.

\begin{figure}[h]
  \centering
  \includegraphics[width=9cm]{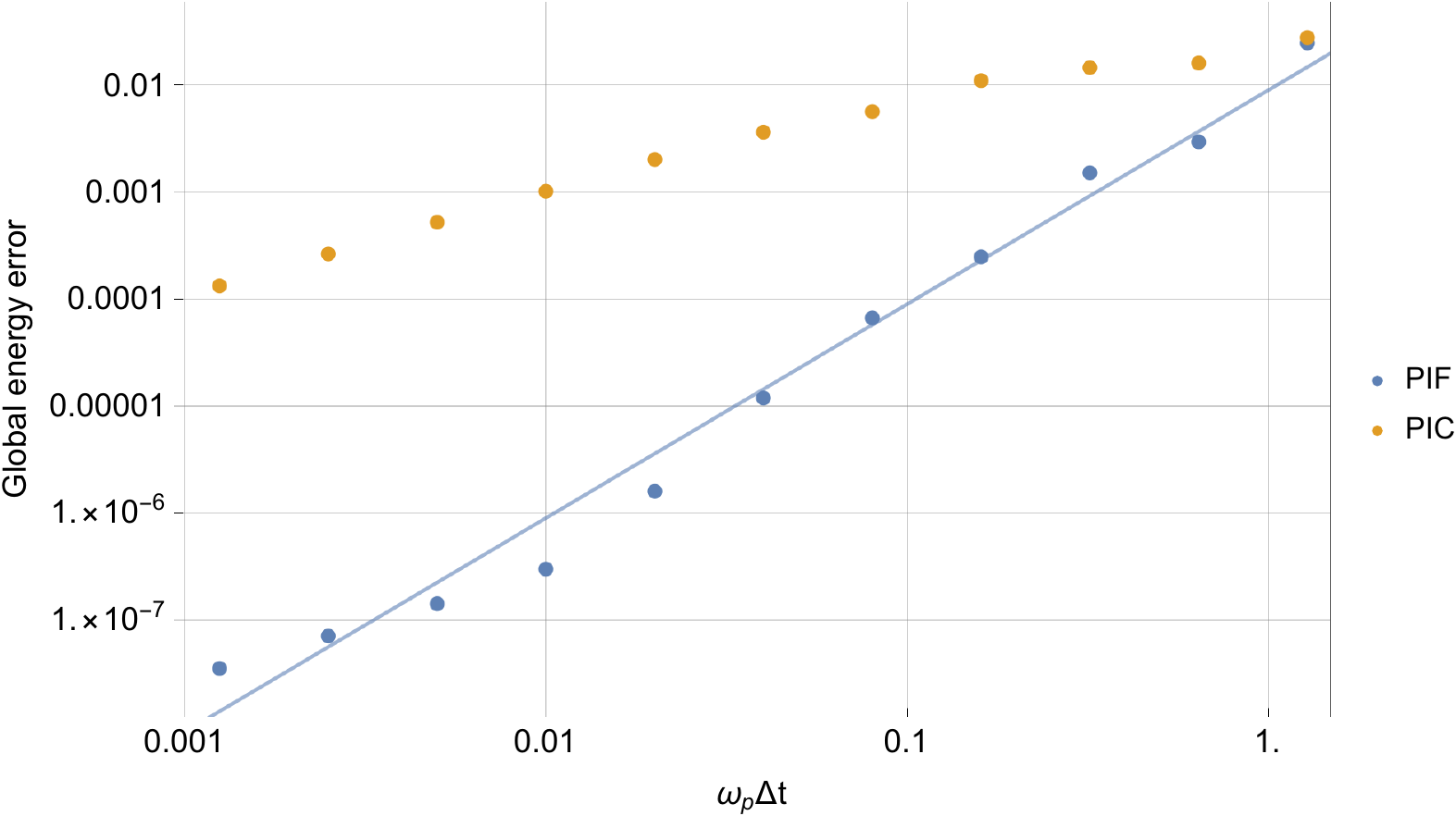}
  \caption{1D relative global error with respect to timestep size for a Landau damping test case
    with 32 grid cells/modes and 20,000 particles. In agreement with
    \cite{evstatiev_variational_2013}, we see global convergence of roughly $O(\Delta t^2)$ for PIF
    in the asymptotic regime.}
  \label{fig:convergence1}
\end{figure}

\begin{figure}[h]
  \centering
  \includegraphics[width=9cm]{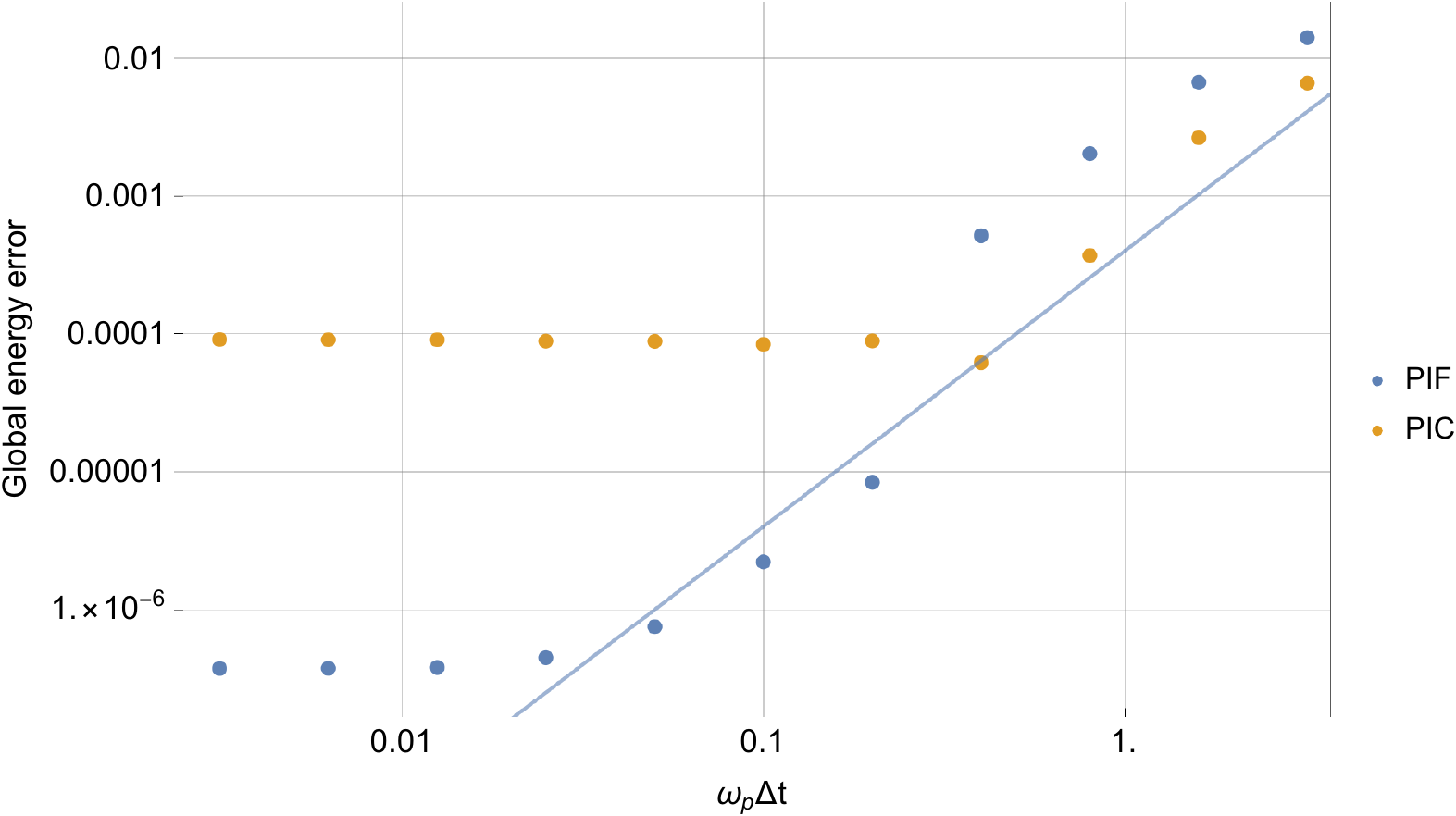}
  \caption{2D global error with respect to timestep size for a Landau damping test case with
      $32 \times 32$ grid cells/modes and 20,000 particles, also in agreement with literature.}
  \label{fig:convergence2}
\end{figure}

We have shown energy conservation in the continuous time limit, but an actual simulation involves
discretization in the time domain. Numerically, we observe that the global truncation timestep error
in the total energy of PIF converges as $O(\Delta t^2)$, in agreement with
\cite{evstatiev_variational_2013} and \cite{huang_finite_2016}. This holds in any number of
dimensions, as illustrated in Fig.\ \ref{fig:convergence1} and Fig.\ \ref{fig:convergence2}.
Note that, although PIC is also quadratically convergent, this is only true for a sufficiently
  fine grid, while PIF displays quadratic convergence independent of the number of modes. For the
  parameters chosen here, we see that PIC and PIF disagree.

\section{Performance and Scaling} \label{sec:perf}

\begin{figure}[h]
  \centering
  \includegraphics[width=10cm]{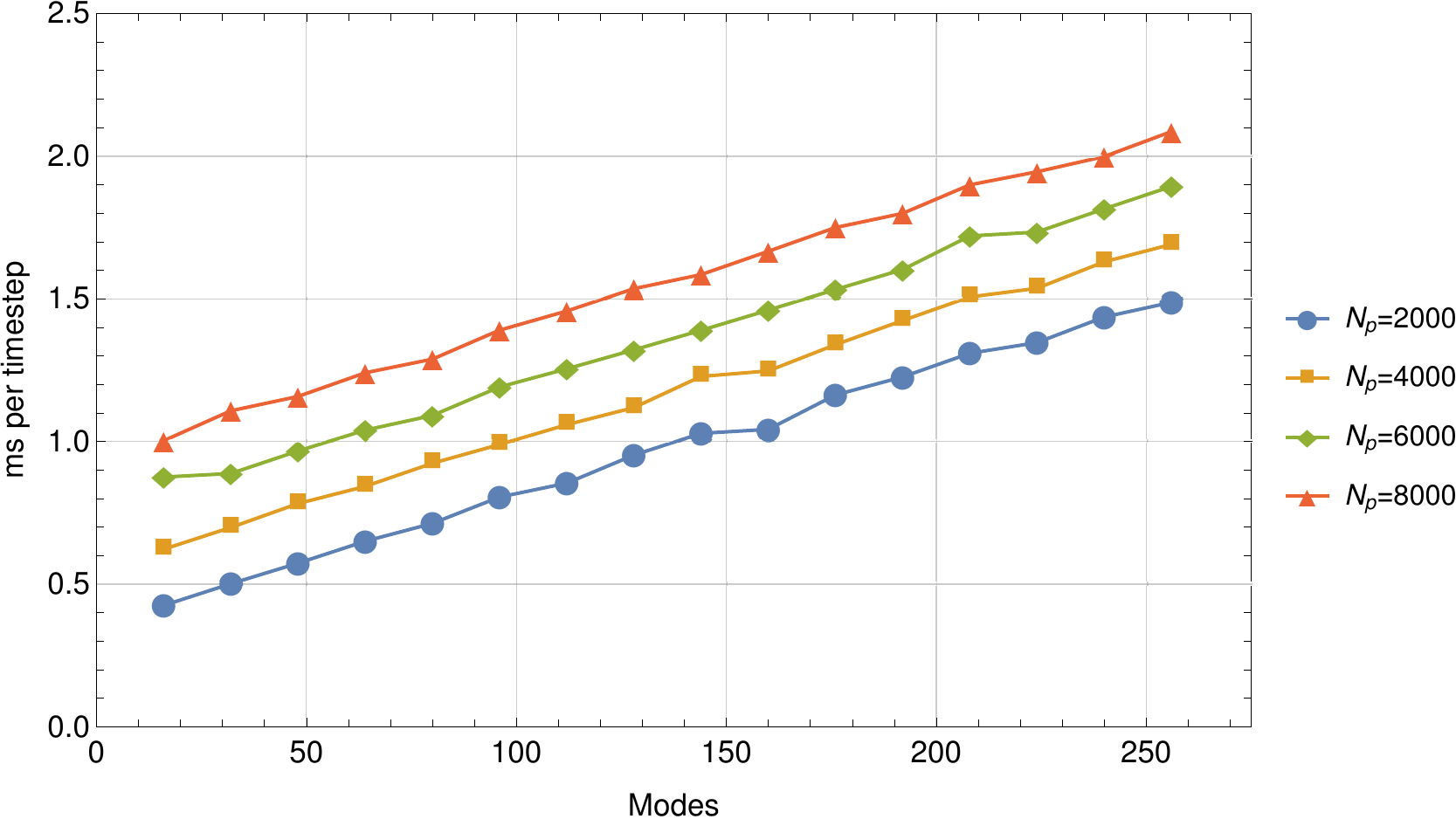}
  \caption{1D PIF performance with respect to particle number and mode number. We see asymptotic time
    complexity approximating $O(N_p + N_m \log N_m)$ as predicted by theory.}
  \label{fig:npng1}

  \includegraphics[width=10cm]{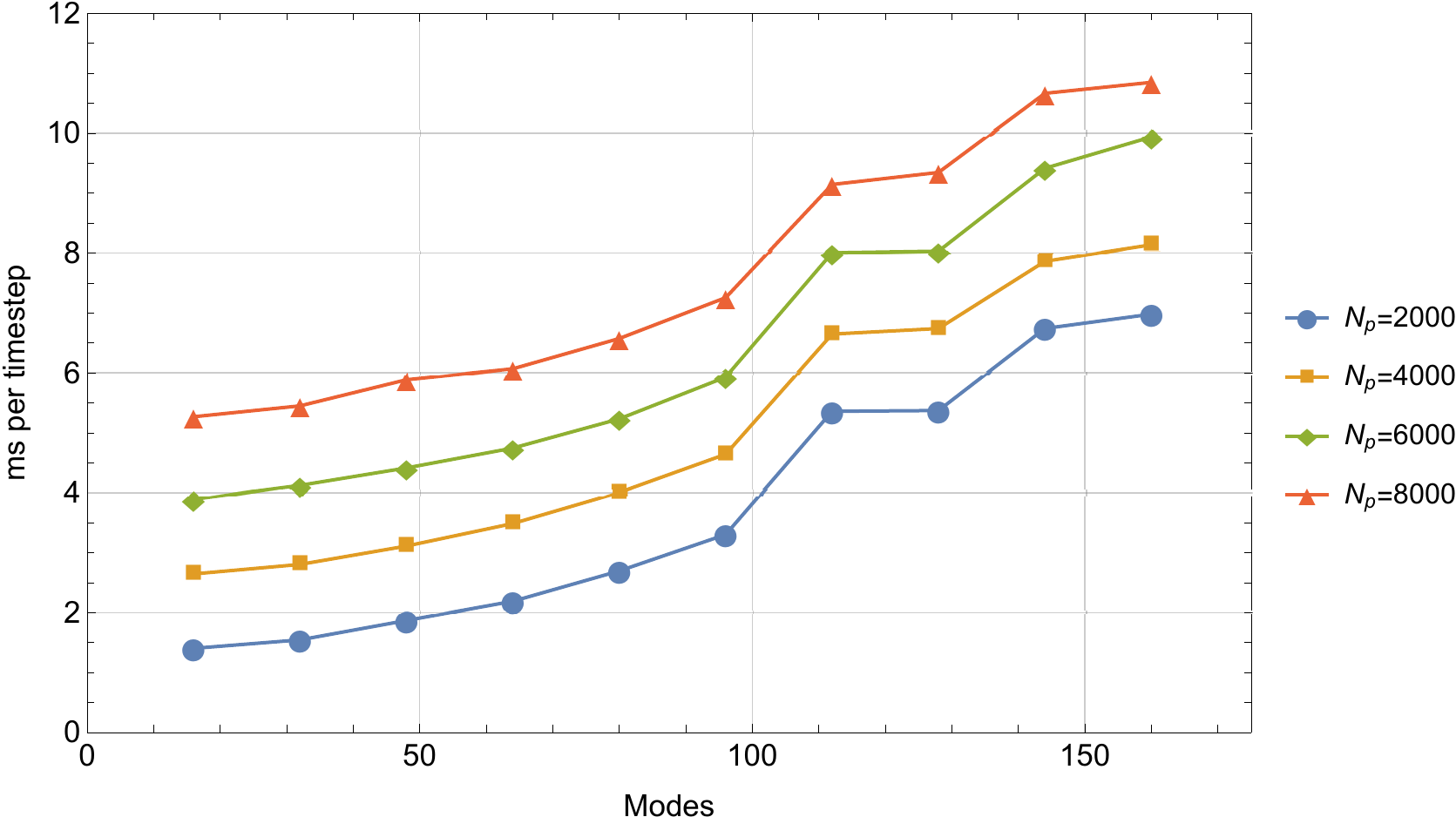}
  \caption{2D PIF performance with respect to particle number and mode number. As in the 1D case, we
    see agreement with the theoretical scaling of $O(N_p + N_m^2 \log N_m^2)$.}
  \label{fig:npng2}
\end{figure}

The asymptotic runtime of the USFFT algorithm is $O(N_p + N_m \log N_m)$ \cite{beylkin_fast_1995}.
Because the field solve and the particle push take $O(N_m)$ and $O(N_p)$ time respectively, the
total asymptotic runtime of the particle-in-Fourier algorithm is $O(N_p + N_m \log N_m)$ for a
single iteration. This is identical to the standard PIC algorithm, assuming the field solve is done
spectrally. This scaling is demonstrated numerically in Fig.\ \ref{fig:npng1} and
\ref{fig:npng2}.

Though we only provide one and two-dimensional implementations, the same scheme can be extended to
three dimensions, while still scaling well. In any number of dimensions, the generalized USFFT
requires $O(N_p + N_m^D \log N_m^D)$ time (where $D$ is the dimensionality of the system). As an
ordinary $D$-dimensional FFT requires $O(N_m^D \log N_m^D)$ time, the scaling is identical to that
of PIC regardless of the dimensionality.

In addition, because convolution with an arbitrary shape function corresponds to multiplication in
Fourier space, increasing the width of the shape function beyond a single grid cell requires no
additional time. This can provide a huge advantage for PIF as higher order shape functions (even
Gaussian) are essentially free, whereas in PIC, higher order shape functions involve many more
calculations and difficult-to-parallelize scatter operations.

For reasonable parameters, the $N_p$ term dominates over the $N_m \log N_m$ term. This suggests the
following parallelization scheme for  PIF (similar to one commonly used
  for PIC): we divide the particles between nodes, with each node performing the field solve for
its own particles (i.e. each node evaluates the serial USFFT independently for a fraction of
  the particles). Before the push step, the E-fields from every node are summed together. To test
this scheme, we created a simple shared memory implementation of it in the one dimensional versions
of PIF and PIC.

\begin{figure}[h]
  \centering
  \includegraphics[width=9cm]{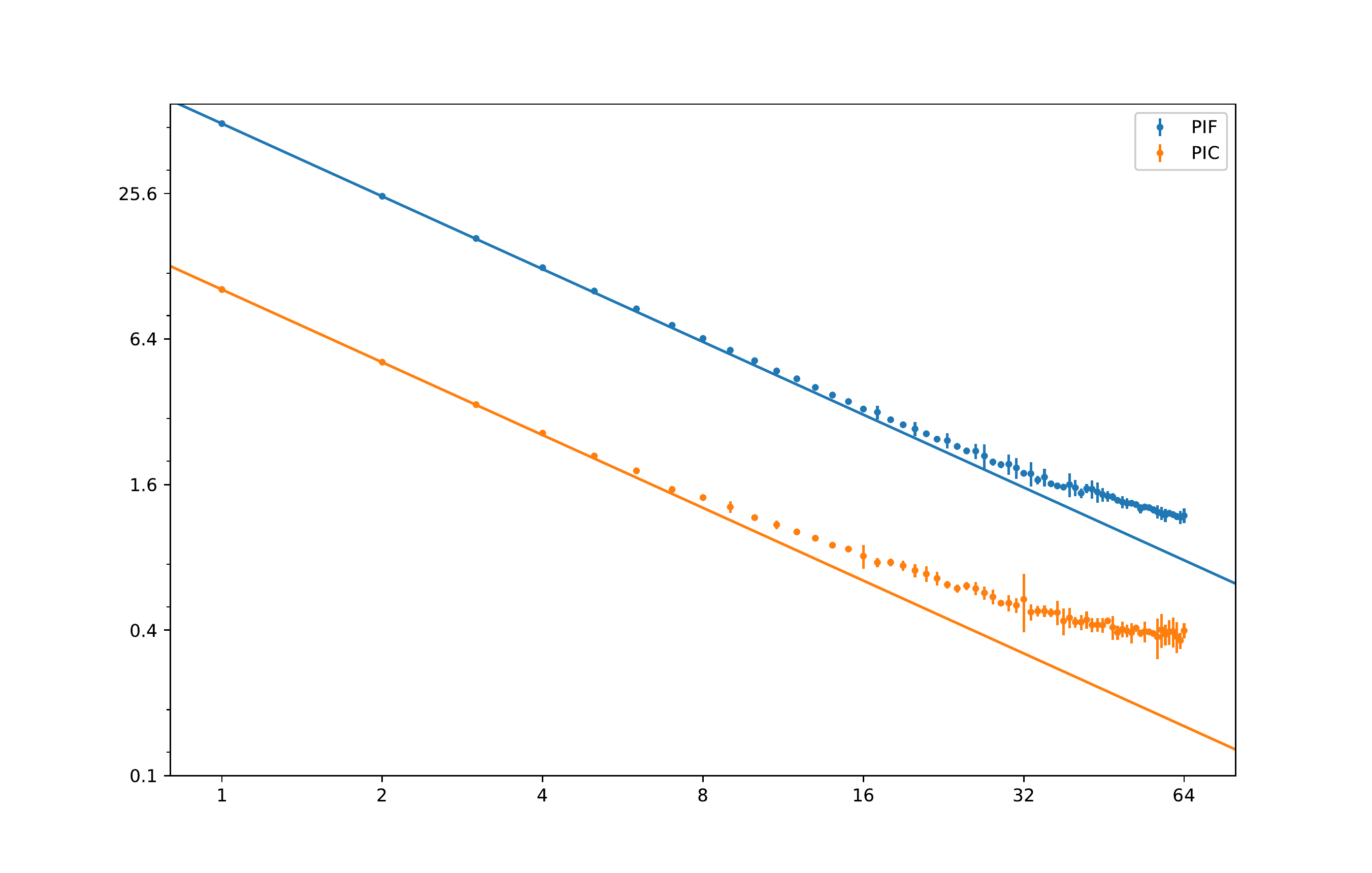}
  \caption{1D PIF has comparable strong scaling to 1D PIC as the number of threads is increased from
    1 to 64, with 80000 particles and 64 grid cells.}
  \label{fig:strongscale}
\end{figure}

In order to verify that this scaling holds, the two 1D codes were run for 1000 time steps
with 80000 particles and 64 grid cells/mode on a single 68-core Xeon Phi node of the Cori
supercomputer at the National Energy Research Scientific Computing Center (NERSC). Both codes
  were compiled with GCC. The number of OpenMP threads was varied from 1 to 64, with the problem
size kept constant. For each data point, the code was run 4 times, with the variation in runtime
indicated by the error bars in Fig. \ref{fig:strongscale}. Both algorithms exhibited similar strong
scaling, as shown in Fig.\ \ref{fig:strongscale}. On average, PIF required 2.9 times as much time
as PIC, and the two algorithms demonstrated comparable scaling.

\begin{figure}[h]
  \centering
  \includegraphics[width=9cm]{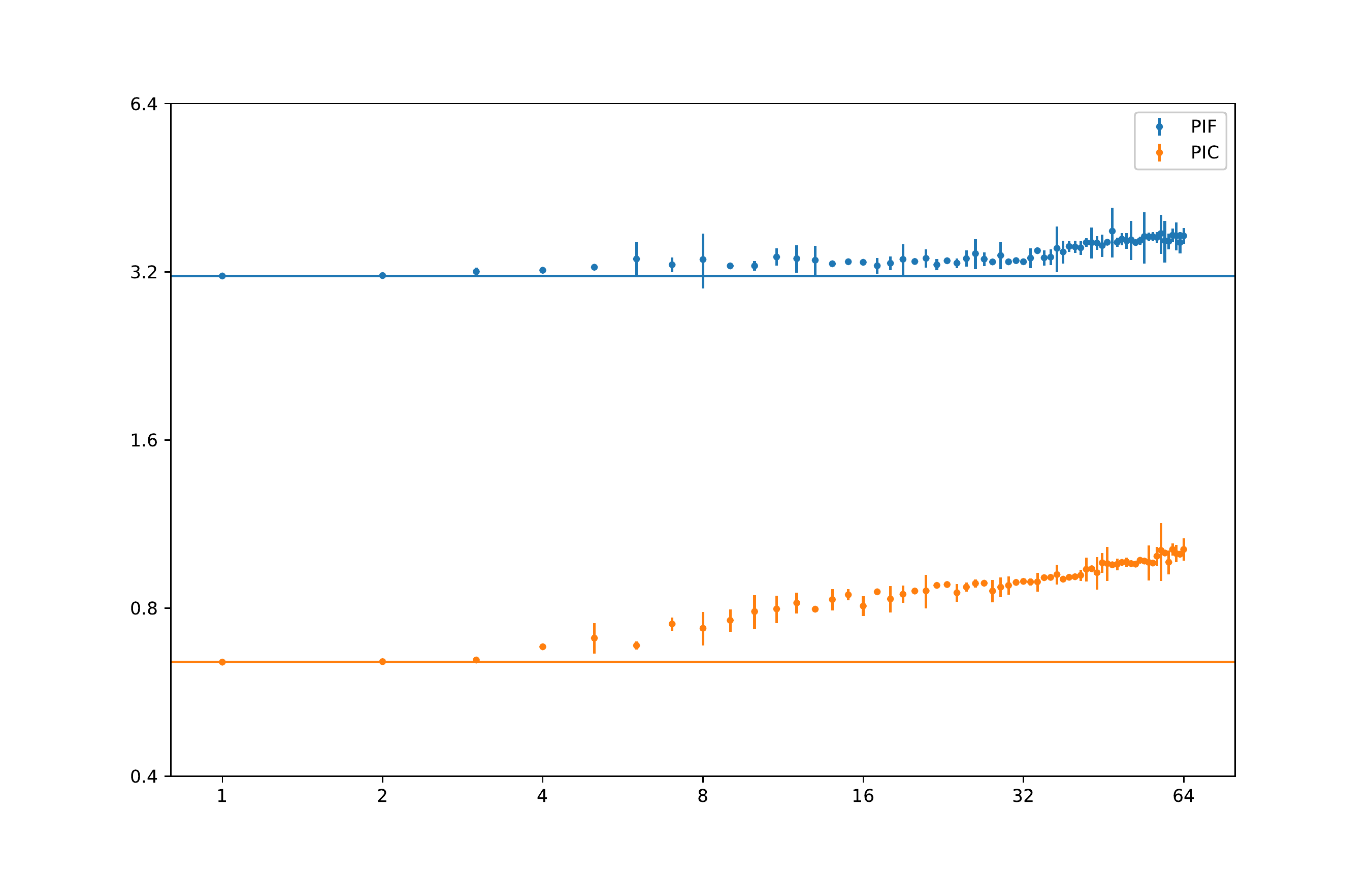}
  \caption{1D PIF has superior weak scaling to 1D PIC as the number of threads is increased from 1 to
    64, with 20000 particles/thread and 64 grid cells.}
  \label{fig:weakscale}
\end{figure}

Weak scaling was investigated by increasing the problem size, such that each run had 20000 particles
per thread. The number of grid cells was kept constant between runs. Results are shown in Fig.\
\ref{fig:weakscale}. We see that PIF demonstrates better weak scaling than our implementation of PIC
for large numbers of particles. We suspect that this is because of the difficulty of parallelizing the
deposition step due to the scatter operations involved.

\section{Discussion} \label{sec:concl}

We have demonstrated that the PIF gridless scheme is a feasible approach to plasma simulation, as it
can be implemented with comparable performance and identical scaling to the conventional PIC method
while conserving both energy and momentum in the continuous-time limit. We have provided an analysis
of these conservation properties and verified them through several numerical experiments. In
addition, we have verified that the results of the PIF model agree with theory for standard test
cases such as Landau damping and the two stream instability.

We did find that PIF was slower than a reference implementation of PIC by a factor of approximately
2.9 in the 1D case and 6.8 in the 2D case. However, PIF demonstrated better weak and strong scaling
by avoiding costly scatter operations during the deposition step. Furthermore, the computational
cost of PIF is independent of the particle shape function, permitting the use of higher order or
Gaussian weighting schemes without additional time.

We expect that the PIF algorithm can be generalized to three dimensional energy conserving
electrostatic and electromagnetic models, as the field solves can still be performed with spectral
methods \cite{birdsall_plasma_1985}. This can be accomplished through the use of three dimensional
USFFTs, still with a computational cost of $O(N_p + N_m^D \log N_m^D)$ \cite{beylkin_fast_1995}. In
addition, we expect that our method could be generalized to $\delta f$ based codes
\cite{parker_fully_1993}, which are of particular interest to fusion plasmas.

In some applications, the simulation model may be periodic in one or more directions, and one could
use a PIF representation in the periodic directions and PIC in the non-periodic ones. For example,
in a toroidal fusion plasma, a Fourier representation in the toroidal direction is appropriate. Such
a mixed basis approach needs to be explored further.

\section*{Acknowledgments}

Research supported by the United States Department of Energy under grants DE-FG02-08ER54954 and
DE-SC0008801.

\bibliographystyle{elsarticle-num}
\bibliography{particle-fourier}

\end{document}